\documentstyle[aps,prl,epsfig,twocolumn]{revtex}

\def\be{\begin{equation}}
\def\ee{\end{equation}}
\def\bea{\begin{eqnarray}}
\def\eea{\end{eqnarray}}
\def\bma{\begin{mathletters}}
\def\ema{\end{mathletters}}

\begin{document}
\draft

\title{Quantum--Classical Correspondence: Controlling Quantum
Transport by State Synthesis in Ion Traps}

\author{Juan F. Poyatos$^1$ and Gonzalo Garc\'{\i}a de
Polavieja$^2$~\cite{bothequal}}

\address{$^1$Centre for Quantum Computation, Clarendon Laboratory,
Parks Road, Oxford OX1 3PU, United Kingdom}
\address{$^2$Physical and Theoretical Chemistry Laboratory, South
Parks Road, Oxford OX1 3QC, United Kingdom}

\date{\today}

\maketitle

\begin{abstract}
A procedure to enhance the quantum--classical correspondence even
in situations far from the classical limit is proposed. It is
based on controlling the quantum transport between classical
regions using the capability to synthesize arbitrary motional
states in ion traps. Quantum barriers and passages to transport
can be created selecting the relevant frequencies. This technique
is then applied to stabilize the quantum motion onto classical
structures or alter the dynamical tunneling in nonintegrable
systems.
\end{abstract}


\narrowtext

%--------------------------------------------------------------------

Systems for which predictions can be made using classical mechanics
can show quantum mechanical properties under suitable experimental
conditions. The observables measured can be of a non-classical class
as in Bell experiments. States can also be chosen to deviate
from classical behavior, an idea that we will exploit in the
present Letter. Despite this, the correspondence principle continues
to be a guide to the study of the interrelation between
quantum and classical mechanics, specially applied
to the study of atomic and molecular systems~\cite{Pe95}.
More recently, a new class of experiments for which Hamiltonians
can be engineered and detailed properties can be monitored have
allowed to apply this correspondence in more detail. Cold atoms
experiments have shown the possibility of inducing quantum dynamics
with a particularly appealing classical limit, the $\delta$-kicked
rotor. These experiments observed dynamical localization, accelerator
modes and only recently the effect of noise and dissipation~\cite{CoAt}.
In addition there have been theoretical
proposals of experimental configurations in ion traps to study
dynamical localization~\cite{DyLo}, revivals~\cite{Br97} and state
sensitivity~\cite{Ga98,Pol98}.

In this Letter we show how to enhance the quantum--classical correspondence
by controlling and monitoring quantum transport in an ion trap. The
control is achieved not only by engineering the Hamiltonian
but most importantly by state synthesis.
To monitor the relevant effects we use tomography and simpler
alternative techniques. Trapped ions advantages to both control
and monitor have already been used for
the study of other aspects of quantum systems, like
the quantum Zeno effect~\cite{It90}, possible nonlinear variants of
quantum theory~\cite{Bo89}, reservoir engineering~\cite{Po96} and
quantum computation~\cite{QuCo}.

To see the relative contribution of the classical backbone
and the purely quantum effects consider the $Q$ distribution
that it is measured in tomography~\cite{Tomo}.
Its continuity equation is of the form
\be
\frac{\partial Q}{\partial t}+\{Q , H\} + Z(V,Q,t,\hbar)=0,
\label{eq:continuity}
\ee
with $Q(\alpha;t)=|\langle\alpha| \Psi(t) \rangle|^{2}$, being
$|\alpha \rangle$ a coherent state, $V$ the potential and $\{,\}$
the classical Poisson bracket. This continuity equation is the
classical Liouville equation plus a quantum term $Z(V,Q,t,\hbar)$
given by
$Z= -\frac{2}{\hbar} \rm{Im} \{ \langle \Psi | \alpha \rangle
V \left( x/2 + i \hbar \partial / \partial p \right) \langle
\alpha | \Psi \rangle \}$.
An ideal experimental setup to study the
quantum--classical correspondence would then have the possibility to control
{\it all} three dependencies of $Z$, i.e.,
the potential, the quantum state and an effective Planck constant.
Moreover, it should be possible to have measurable quantities
showing the relative importance of classical and quantum
contributions, e.g. the $Q$ distribution itself.
All of the above conditions can be fulfilled in the case of an
harmonically trapped ion. This is possible due to the interaction
between the internal (electronic) and external (vibrational) degrees
of freedom of the ion by means of laser pulses
in resonant and non--resonant regimes~\cite{Iotr}.

In the quantum--classical boundary both the classical Liouville
and the quantum terms in the continuity equation in~(\ref{eq:continuity})
are relevant. The quantum contribution to the transport is
state dependent and the final flow is diverted from the classical flow
by an amount that depends on the quantum state.
We first pick out the classical backbone structure
and then study how to manipulate the quantum transport
depending on the synthesized state. We need to construct
a family of Hamiltonians that have regular, mixed or chaotic phase
space. In the trapped ion setup we make use of the harmonic delta
kicked Hamiltonian, first introduced in~\cite{Ga98},
describing a harmonic oscillator periodically perturbed by
non--linear position dependent delta kicks.
We consider a single trapped ion in a harmonic potential~\cite{Iotr}
with two internal levels $|e\rangle$ and $|g\rangle$ with
transition frequency $\omega_0$ interacting with a time dependent
laser pulse of near--resonant light of frequency $\omega_L$ which is
rapidly and periodically switched.
For sufficiently large detuning $\delta=\omega_0-\omega_L$ the
excited state amplitude can be adiabatically eliminated.
The Hamiltonian in this limit is given by~\cite{Ga98}
\be
\label{eq:Ham}
H=H_0+K [\cos(2k_Lx)+1] |g\rangle \langle g|
\sum^\infty_{n=-\infty} \delta (t-n\tau),
\ee
where $H_0$ is the harmonic oscillator Hamiltonian,
$K$ is the kick strength related to the laser beam power,
$k_L=2\pi/\lambda$ the laser wave number, and
$\tau$ the time between kicks.
A particularly important parameter is the so--called
Lamb--Dicke parameter $\eta=k_L\sqrt{\hbar/2m\nu}$, where $\nu$
is the oscillator frequency and $m$ the mass. Thus by varying
$\eta$ it is possible to change the {\it effective} $\hbar$ of
the system, i.e. doing it more classical.

In the following the relevance of the classical backbone is shown by
predicting the time averaged quantum $Q$ function by means of
classical theorems. We write the potential for a phase space region
as $V=V_{0} + V_{1}$, with $V_{0}$ and $V_{1}$ the unperturbed an
perturbed potentials that in general have different phase space
topologies. The time averaged dynamics of an initially synthesized
motional Fock state is shown to be predicted from knowledge of the
classical solution for $V_{0}$ and the Kolmogorov-Arnold-Moser (KAM)
and the Poincare-Birkoff (PB) theorems~\cite{Tabor}. These two theorems
in conjunction mean that, increasing the perturbation $V_1$, the phase
space tori break in increasing irrationality values of the ratio of
their winding numbers $r=\omega_{1}/\omega_{2}$ with
$\omega_1$ and $\omega_2$ the two
frequencies of a given torus~\cite{Note0}. When the winding number
is sufficiently close to the rational number $l/s$ the torus breaks
into alternative $ks$ stable and unstable points,
with $k,l,s$ integers~\cite{Tabor}.

%%%%%%%%%%%%%%%%%%%%%%%%%%%%
% Figure 1
\begin{figure}[tbp]
\epsfig{file=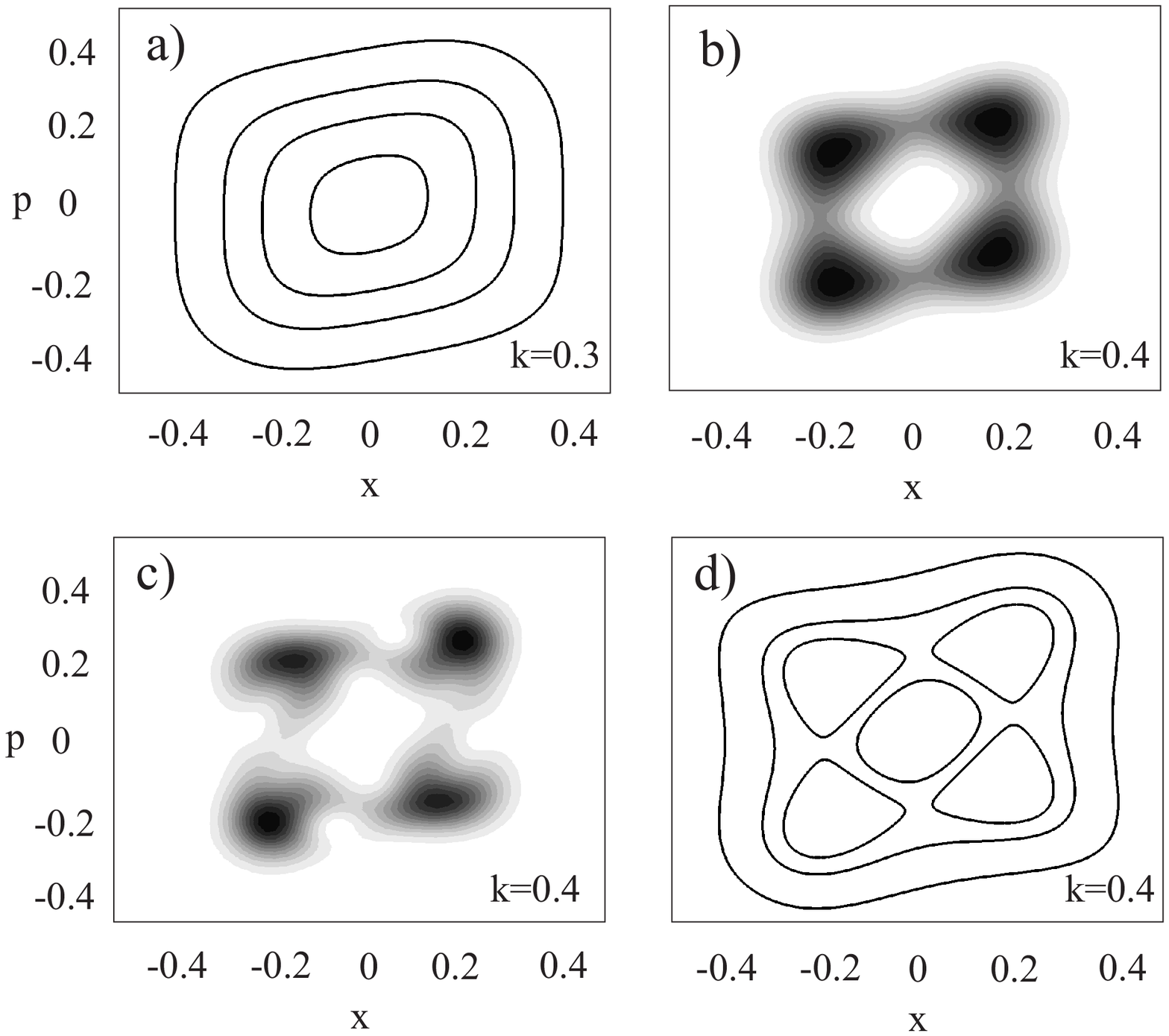,width=8cm}
\caption{
{\sf
(a) Stroboscopic classical map with $k=0.3$. The
rational number to which the winding number approaches is 1/4
for the initial condition (x,p)=(0.22,0).
(b) $Q_T$ function with $k=0.4$ and $\eta=0.25$ (see text).
(c) $Q$ function averaged over three different tomographic
measurements after applying 21, 22 and 23 kicks. All parameters
as before.
(d) Stroboscopic classical map with $k=0.4$.
$\nu\tau=1.8$ for all the cases.
$x$ and $p$ are in units of $\lambda$ and $m\nu\lambda$ in
all discussions and $k=\sqrt{2}K\eta^2/\hbar$.
}
}
\end{figure}
%%%%%%%%%%%%%%%%%%%%%%%%%%%%

Fig. 1(a) shows the classical
stroboscopic map for $V_0$. Following the values of the winding
number increasing the perturbation we determine which torus is
breaking first. We thus predict the averaged quantum behavior
with total potential $V=V_0 + V_1$ using only the classical information
from the values of the winding numbers with $V=V_0$ and the
classical theorems.
Then a $|n\rangle=|7\rangle$ Fock state
is predicted to have an averaged $Q$ function with four
maxima (minima) on the classically stable (unstable) points
when increasing the perturbation of the system.
Fig. 1(b) shows the time-averaged Q function
\be
Q_T=\lim_{M \rightarrow \infty} \frac{1}{M \tau}
\sum_{m=0}^{M-1} Q(m \tau)=\sum_{\mu} |\langle \alpha|\mu \rangle|^{2}
|\langle \Psi|\mu \rangle|^{2} ,
\label{eq:qaverage}
\ee
with $|\mu\rangle$ the Floquet states fulfilling
${\hat U}(\tau) |\mu \rangle = e^{-i \epsilon_{\mu}} |\mu \rangle$,
where ${\hat U}(\tau)$ is the time evolution operator referring to one
period $\tau$ and $\Psi$ is a given initial state.
The four predicted maxima are clear in this Figure.
Fig. 1(c) displays the $Q$ function averaged at three
consecutive times, as it could be measured experimentally using
three tomographic measurements~\cite{Tomo}
showing the same structure~\cite{Notea0}.
The classical stroboscopic map for the perturbed
case in Fig. 1(d) shows clearly that $Q_T$ reveals the
classical backbone in many details.
Any other Fock state prepared would scan in the same way
different structures of this or alternative classical maps.

We now propose to go a step further in the study of the
quantum--classical correspondence controlling the relevance of the
quantum and classical terms in~(\ref{eq:continuity}) by state
synthesis. We want to prepare a state initially localized in a
classical region $A$ with a {\it barrier} or a {\it passage} for
transport to a different classical region $B$. Take as a starting
state a coherent state localized on $A$, $\phi_{A}(0)$. Its
averaged transport to region $B$, represented by another state
$\phi_{B}$, is given by $P(|\phi_{A}\rangle,|\phi_{B}\rangle)=
\lim_{M \rightarrow \infty} \frac{1}{M \tau} \sum_{m=0}^{M-1}|
\langle \phi_{B}| \phi_{A} (m \tau)\rangle|^{2}$, that we can
write as
$P(|\phi_{A}\rangle,|\phi_{B}\rangle)=\sum_{\mu}|\langle\phi_{B}|
\mu\rangle|^{2}|\langle\phi_{A}(0)|\mu\rangle|^{2}$ using the
Floquet basis. For $P(|\phi_{A}\rangle,|\phi_{B}\rangle)$ not to
be zero, the states $\phi_{A}(0)$ and $\phi_{B}$ must have nonzero
overlap with common Floquet states denoted as $\{ |\mu\rangle_{A
\cap B} \}$. To form a new state that minimizes,
$\varphi_{A'}^{-B}$, or maximizes, $\varphi_{A'}^{+B}$, the
transport to region $B$ we eliminate (enhance) from $\phi_{A}(0)$
the Floquet components $\{\mu_{A \cap B}\}$ as
\be
|\varphi_{A'}^{\pm B} \rangle =N_1 ( |\phi_{A} (0)\rangle
 \pm \sum_{\mu_{A \cap B}} r_{\mu} |\mu \rangle \langle \mu|
\phi_{A}(0) \rangle),
\label{eq:transport1}
\ee
with $r_{\mu}=0$ or $1$ when the corresponding weight $|\langle
\mu| \phi_{A} \rangle|^{2}$ is smaller or greater than a value
$c_{\rm tol}$, respectively and $N_i$ will be normalization
constants from now on. The value  $c_{\rm tol}$ is chosen to be
the minimum possible subject to the condition that the new region
$A'$ is sufficiently close to the initial region of localization
$A$. We propose then the synthesis of the states
\be
|\Psi_{A'}^{\pm B} \rangle =N_2 \sum_{n=0}^{n_{\rm exp}}
| n \rangle \langle n | \varphi_{A'}^{\pm B} \rangle,
\label{eq:transport2}
\ee
with $\{|n\rangle\}$ the Fock basis and where we consider only an
experimentally feasible maximum number of Fock states $n_{\rm
exp}$~\cite{Note1} and amplitudes $\langle n|\varphi_{A'} ^{\pm
B}\rangle$ chosen to approximate the theoretical
state~(\ref{eq:transport1}).

%%%%%%%%%%%%%%%%%%%%%%%%%%%%
% Figure 2
\begin{figure}[tbp]
\epsfig{file=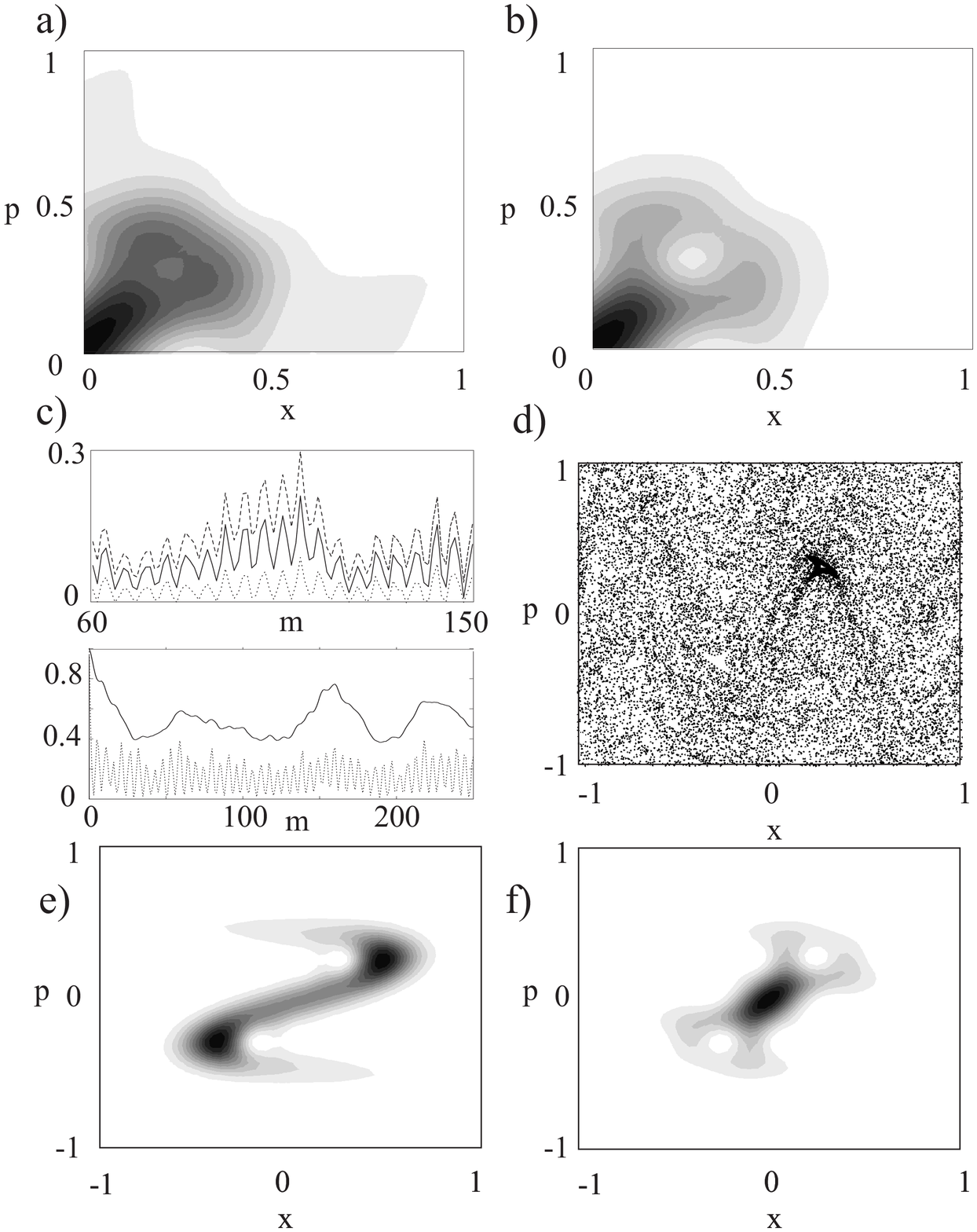,width=8cm}
\caption{
{\sf
(a) $Q_T$ function for $\phi_A$ coherent state.
(b) $Q_T$ function for modified $\Psi_{A'} ^{- B}$ state.
(c) Upper Plot. Correlation function for $\phi_A$,
$\Psi_{A'} ^{- B}$ (dot) and $\Psi_{A'} ^{+ B}$ (dashed).
Lower Plot. Autocorrelation function for $\Psi_{A'} ^{\rm sta}$ state
and $\phi_A$ coherent state (dot).
(d) Stroboscopic classical map.
(e) $Q$ for a initial coherent state centered at $(x_A,p_A)=(0,0)$ after two kicks.
(f) As before for a $\Psi_{A'} ^{\rm sta}$ state.
All figures obtained with the parameters: $k=2.7$, $\nu\tau=1.7$
and $\eta=.5$. See text for more details of each case.
}
}
\end{figure}
%%%%%%%%%%%%%%%%%%%%%%%%%%%%

The enhanced quantum--classical correspondence is thus obtained
even {\it far} from the classical limit by choosing classical
regions $A$ and $B$ and controlling the transport between them as
described above. Some comments for this construction are
necessary: (a) We can give up the condition that the initial state
must be localized in region $A'$ slightly bigger than the region
$A$ of the minimum uncertainty wavepacket. For some experiments it
is interesting to have a larger variance of the initial state or
an additional localization in a different region $A''$. (b) If we
only want to modify the transport to region $B$ and not to other
regions $C$ we have to use $r_{\mu} =0$ in~(\ref{eq:transport1})
for the Floquet states that have important overlap with a state
$\phi_{C}$ centered on $C$. (c) The states localized in the
regions $A$, $B$ and $C$ do not need in general to be minimum
uncertainty states but we have the freedom to use any states
related to these phase space regions.

We now discuss first a general example of modified transport
properties of a given state and we then study stabilization, and
modification of dynamical tunneling as two interesting examples.
We want to synthesize a state $\Psi_{A'} ^{-B}$ initially
localized on $(x_A,p_A)=(0.1,0.2)$ that avoids transport to
$(x_B,p_B)=(0.3,0.3)$. We use coherent states for $\phi_A$ and
$\phi_B$ and a value $c_{\rm tol}=0.001$ so all the Floquet states
with important overlap with $\phi_{B}$ have $r_{\mu}=1$
in~(\ref{eq:transport1}). An experimentally realizable state with
$n_{\rm exp}=10$ can be constructed with overlap $|\langle
\Psi_{A'} ^{-B}|\varphi_{A'} ^{- B} \rangle |^2=0.9$ to the
theoretical state. Despite both wavefunctions
$\langle\alpha|\phi_{A}\rangle$ and $\langle\alpha|\Psi_{A'}^{-B}
\rangle$ being initially localized around $A$, with
$|\langle\phi_{A}|\Psi_{A'}^{-B}\rangle|^2=0.82$, it is clear from
Fig. 2(a) and (b) that their $Q_T$ functions are very different.
The $Q_T$ function for $\Psi_{A'} ^{- B}$ shows a clear hole at
the location $B$, in contrast to the minimum uncertainty state
$\phi_{A}$. The correlation function $P(m \tau)=|\langle \phi_{B}
| \hat{U}^{m}(\tau)| \Phi \rangle|^{2}$ is shown in Fig. 2(c) for
$\Phi=\phi_{A}$ and $\Psi_{A'} ^{- B}$, the latter showing a
significant decrease. These correlation functions can be measured
experimentally by applying a displacement associated to $\phi_B$
and then measuring the Fock ground state population both steps
possible to implement in a trap~\cite{Po96b}. Successful results
can also be obtained for the state $\Psi_{A'}^{+B}$ that presents
a maximum on region $B$ in $Q_T$, see Fig. 2(c).

Particular instances of the states $\Psi_{A'}^{\pm B}$ are of
special relevance. The stabilization of the vibration of the ion
can be understood as a modified transport problem. In this case
$A$ is the region of localization and $B$ the rest of the phase
space and we are interested in constructing a state $\Psi_{A'} ^{-
B} $ that we will name for the stabilization case as $\Psi_{A'}
^{\rm sta} $. An alternative way to understand
expression~(\ref{eq:transport2}) for this situation now in terms
of the dynamics is given in the following. The autocorrelation
 function of a typical state
$|\langle \phi_A(0)|\phi_A(t)\rangle|^2$ will show an initial
maximum (however small) at $t_M>0$. We can now clean the state
$\phi_{A}$ of the Floquet components that do not contribute to
this maximum and therefore create a new state stabilized on $A$.
Note first that a particular Floquet state $|\mu_{0}\rangle$ can
be obtained from the time-dependent vector $|\phi_{A}
(t)\rangle=\sum_{\mu} |\mu \rangle \langle\mu|\phi_{A}(0) \rangle
\exp(-i \epsilon_{\mu} t )$ (a solution of the Schr\"odinger
equation {\it only} for $t=m \tau$) as $|\mu_0\rangle \propto
\lim_{T \rightarrow \infty} G_{T, \mu_0}$ with $G_{T, \mu_0}
\equiv \int_{-T}^{T} dt  \ |\phi_{A} (t) \rangle \exp(i
\epsilon_{\mu_{0}} t)$. A state related to the short term dynamics
is then $G_{t_{M},\omega_{0}}$ with $\omega_{0}$ the value of
$\omega$  that makes $S(\omega)=\int_{-t_{M}} ^{t_{M}} dt \langle
\phi_{A}(0)|\phi_{A}(t) \rangle \exp (i \omega t)$ a
maximum~\cite{Pol94}. This state $G_{t_{M}, \omega_{0}}$ can then
be approximated as
\be
\label{eq:Flo}
|\varphi_{A'}^{\rm sta} \rangle =N_{3} \sum_{\mu_{\rm sta}} |\mu
\rangle \langle \mu | \phi_{A} (0) \rangle ,
\ee
with $\mu_{\rm sta}$ the Floquet eigenfrequencies in the interval
$ \omega_{0} -\frac{\pi}{t_{M}} < \omega <
\omega_{0} +\frac{\pi}{t_{M}} $ with a weight $|\langle \mu |
\phi_{A} (0) \rangle|^{2} > a_{\rm sta}$.
The value of $a_{\rm sta}$ is chosen maximum with the requirement
than the state  $ \varphi_{A'}^{\rm sta}$ has a tolerable localization
around $A$. The state $\Psi_{A'} ^ {\rm sta}$ in~(\ref{eq:transport2})
synthesized to approximate $\varphi_{A'}^{\rm sta}$ will then show
an initial localization around $A$
because it selects the short term dynamics and
will recur continuously to $A$ because it is made of
very few selected Floquet states. In fact there are several
maxima in $S(\omega)$ and we can choose the value $\omega_{0}$
with minimum number of Floquet states
maximizing stabilization in this way.

Using this stabilization procedure we have found enhanced
localization onto KAM tori, islands of size smaller than the
effective $\hbar$, cantori or unstable periodic orbits
effects~\cite{Po99}. The following  example of enhanced
localization onto a classical unstable orbit can be realized
experimentally. We consider the stroboscopic map in Fig.~2(d).
Fig.~2(e) shows the $Q$ function for a initial minimum uncertainty
state, $\phi_A$, after two kicks centered at $(x_A,p_A)=(0,0)$.
The state is spread from an unstable periodic orbit along the
unstable manifold. The stabilization is achieved in this case with
a state $\Psi_{A'} ^{\rm sta}$ centered on the middle of the
chaotic region at $(0,0)$ with $|\langle\varphi^{\rm
sta}_{A'}|\Psi^{\rm sta}_{A'}\rangle|^2= 0.72$ and $n_{\rm
exp}=12$, a single Floquet state. The stabilization achieved is
clearly seen in the measurable autocorrelation
function~\cite{Po96b}, see Fig. 2(c) and Fig.~2(f) for the $Q$
function of such stabilized state. This example constitutes a
realizable experiment to directly observe a {\it quantum
scar.}~\cite{He84}

%%%%%%%%%%%%%%%%%%%%%%%%%%%%
% Figure 3
\begin{figure}[tbp]
\epsfig{file=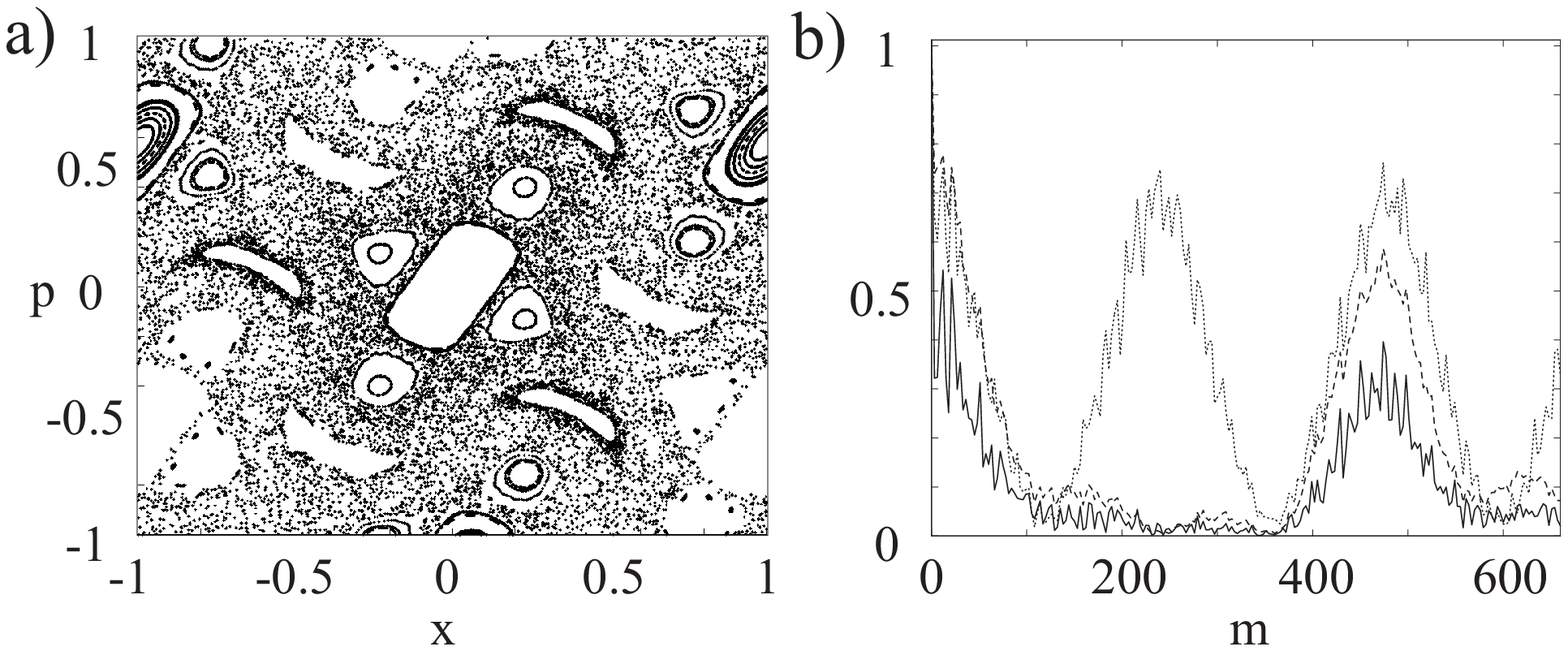,width=8cm} \caption{ {\sf (a) Stroboscopic
classical map with $k=1.2$ and $\nu\tau=2\pi/3$. (b)
Autocorrelation function for an initial coherent state, Floquet
doublet (dot), and $|\Psi^B_{A'}\rangle$ state (dash). Parameters
as before with $\eta=0.5$. } }
\end{figure}
%%%%%%%%%%%%%%%%%%%%%%%%%%%%

As a final example we show how to increase the dynamical tunneling
associated to classically forbidden regions, see~Fig. 3(a). An
initial minimum uncertainty state located at $(x_A,p_A)=(-0.6,0)$
has dynamical tunneling and contributions from a unstable periodic
orbit located at $(x_B,p_B)=(-0.5,0.3)$. A
state~(\ref{eq:transport1}) eliminating the contributions of $B$
is mainly composed of three Floquet doublets reflecting the
oscillations due to quantum tunneling. The autocorrelation
function is shown in~Fig. 3(b) for a state~(\ref{eq:transport2})
with $|\langle\varphi^B_{A'}|\Psi^B_{A'}\rangle|^2=0.85$. This can
be measured by applying displacements associated to the initial
regions of localization or more directly by inverting the unitary
process which created the initial state. An example of
stabilization in this case is choosing a state made of a single
doublet where $|\langle\varphi^{\rm sta}_{A'}|\Psi^{\rm
sta}_{A'}\rangle|^2=0.8$. In this case the oscillations due to the
tunneling are more clearly reflected. Both states with $n_{\rm
exp}=25$ since their location in phase space implies higher Fock
state contributions than previous cases. Choosing different
doublets would give rise to different tunneling times. Finally
notice that by increasing $n_{\rm exp}$, i.e. the overlap to the
theoretical state, all features discussed in previous examples
will be more dramatically shown~\cite{Note1}.

In conclusion, we have presented a way to enhance the
quantum--classical correspondence by modifying quantum transport
in an ion trap. We have specially made use of the ability to
synthesize arbitrary states of motion in a ion trap.

G.G.P. acknowledges a Marie Curie Fellowship. J.F.P. is supported
by the European TMR network ERB-4061-PL-95-1412 and specially
dedicates this paper to M.B.

%%%%%%%%%%%%%%%%%%%%%%%%%%%%%%%%%%%%


\begin{references}
\bibitem{Notea0}
Simulations have been calculated in a truncated Fock basis.
Convergence was verified by variation of the cutoff.
\end{references}
\end{document}